\documentclass[pdflatex,sn-mathphys]{sn-jnl}

\jyear{2022}%

\theoremstyle{thmstyleone}%
\theoremstyle{thmstyletwo}%

\theoremstyle{thmstylethree}%

\usepackage{subfigure}

\begin{document}

\title[Beam test results of 25 $\mu$m and 35 $\mu$m thick FBK UFSD]{Beam test results of 25 $\mu$m and 35 $\mu$m thick FBK ultra fast silicon detectors}

\author*[1]{\fnm{F.} \sur{Carnesecchi}}\email{francesca.carnesecchi@cern.ch}
\author*[2]{\fnm{S.} \sur{Strazzi}}\email{sofia.strazzi2@unibo.it}
\author[2]{\fnm{A.} \sur{Alici}}
\author[4,6]{\fnm{R.} \sur{Arcidiacono}}
\author[7,9]{\fnm{G.} \sur{Borghi}}
\author[7,9]{\fnm{M.} \sur{Boscardin}}
\author[4]{\fnm{N.} \sur{Cartiglia}}
\author[7,9]{\fnm{M.} \sur{Centis Vignali}}
\author[3]{\fnm{D.} \sur{Cavazza}}
\author[8,9]{\fnm{G.-F.} \sur{Dalla Betta}}
\author[5]{\fnm{S.} \sur{Durando}}
\author[4]{\fnm{M.} \sur{Ferrero}}
\author[7,9]{\fnm{F.} \sur{Ficorella}}
\author[7,9]{\fnm{O.} \sur{Hammad Ali}}
\author[4]{\fnm{M.} \sur{Mandurrino}}
\author[3]{\fnm{A.} \sur{Margotti}}
\author[4,5]{\fnm{L.} \sur{Menzio}}
\author[3]{\fnm{R.} \sur{Nania}}
\author[8,9]{\fnm{L.} \sur{Pancheri}}
\author[7,9]{\fnm{G.} \sur{Paternoster}}
\author[2]{\fnm{G.} \sur{Scioli}}
\author[4]{\fnm{F.} \sur{Siviero}}
\author[4,5]{\fnm{V.} \sur{Sola}}
\author[4,5]{\fnm{M.} \sur{Tornago}}
\author[2]{\fnm{G.} \sur{Vignola}}

\affil[1]{\orgname{CERN}, \orgaddress{\city{Geneva}, \country{Switzerland}}}
\affil[2]{\orgdiv{Dipartimento Fisica e Astronomia dell’Università}, \orgaddress{\city{Bologna}, \country{Italy}}}
\affil[3]{\orgname{INFN}, \orgaddress{\city{Bologna}, \country{Italy}}}
\affil[4]{\orgname{INFN}, \orgaddress{\city{Torino}, \country{Italy}}}
\affil[5]{\orgdiv{Università degli Studi di Torino}, \orgaddress{\city{Torino}, \country{Italy}}}
\affil[6]{\orgdiv{Università del Piemonte Orientale}, \orgaddress{\city{Novara}, \country{Italy}}}
\affil[7]{\orgname{Fondazione Bruno Kessler}, \orgaddress{\city{Trento}, \country{Italy}}}
\affil[8]{\orgdiv{Università di Trento}, \orgaddress{\city{Povo}, \country{Italy}}}
\affil[9]{\orgname{TIFPA}, \orgaddress{\city{Trento}, \country{Italy}}}


\abstract{This paper presents the measurements on first very thin Ultra Fast Silicon Detectors (UFSDs) produced by Fondazione Bruno Kessler; the data have been collected in a beam test setup at the CERN PS, using beam with a momentum of 12 GeV/c.
UFSDs with a nominal thickness of 25 $\mu$m and 35 $\mu$m and an area of 1 $\times$ 1 $\text{mm}^2$ have been considered, together with an additional HPK 50-$\mu$m thick sensor, taken as reference.
Their timing performances have been studied as a function of the applied voltage and gain. A time resolution of about 25 ps and of 22 ps at a voltage of 120 V and 240 V has been obtained for the 25 and 35 $\mu$m thick UFSDs, respectively.}




\maketitle

\section{Introduction}\label{sec:intro}

The Ultra Fast Silicon Detectors (UFSDs) \cite{2017Sadrozinski} is a  silicon detector developed to detect charged particles, based on Low Gain Avalanche Diode (LGAD) technology\cite{2014PELLEGRINI}  and improved to achieve excellent time resolution. 
The core of these detectors is a controlled gain layer in addition to a standard n-on-p silicon detector, right below the n-p junction.
Thanks to the moderate internal gain (between 10-70), the Signal-to-Noise ratio (S/N) is improved: for a 45-$\mu$m thick UFSD pixel and a gain of 20-30 a time resolution of $\sim$30 ps \cite{2017Cartiglia} in a beam test setup has been achieved.\\
Thanks to the excellent time resolution, this technology is already envisioned for the detector upgrades at the HL-LHC (both in ATLAS~\cite{atlas} and CMS~\cite{cms}) for 2026. 
However, new generation of compact experiments will require an even better time resolution on a single layer, thus asking for improvements of the timing capabilities of such detectors. A thinner UFSD design may match the requirements \cite{2017Sadrozinski}. 
In this paper, UFSDs with a nominal thickness of 25 $\mu$m and 35 $\mu$m have been considered; their time resolutions have been studied in a beam test setup at PS, CERN. 
Furthermore, 
a 50 $\mu$m thick detector, whose time resolution is known from the literature, has been tested as well as a reference.

The paper is organized as follow: after a description of the detectors, the test beam setup and electronics in section~\ref{sec:setup}, in section~\ref{sec:results} the data analysis and the main results will be described. 


\section{Experimental setup} 
\label{sec:setup}

\subsection{Detectors} 
\label{sec:det}

\label{sec:detectors}
\begin{figure}[htbp]
        \centering%
        \subfigure[\label{fig:2535L}]%
          {\includegraphics [height=3.cm] {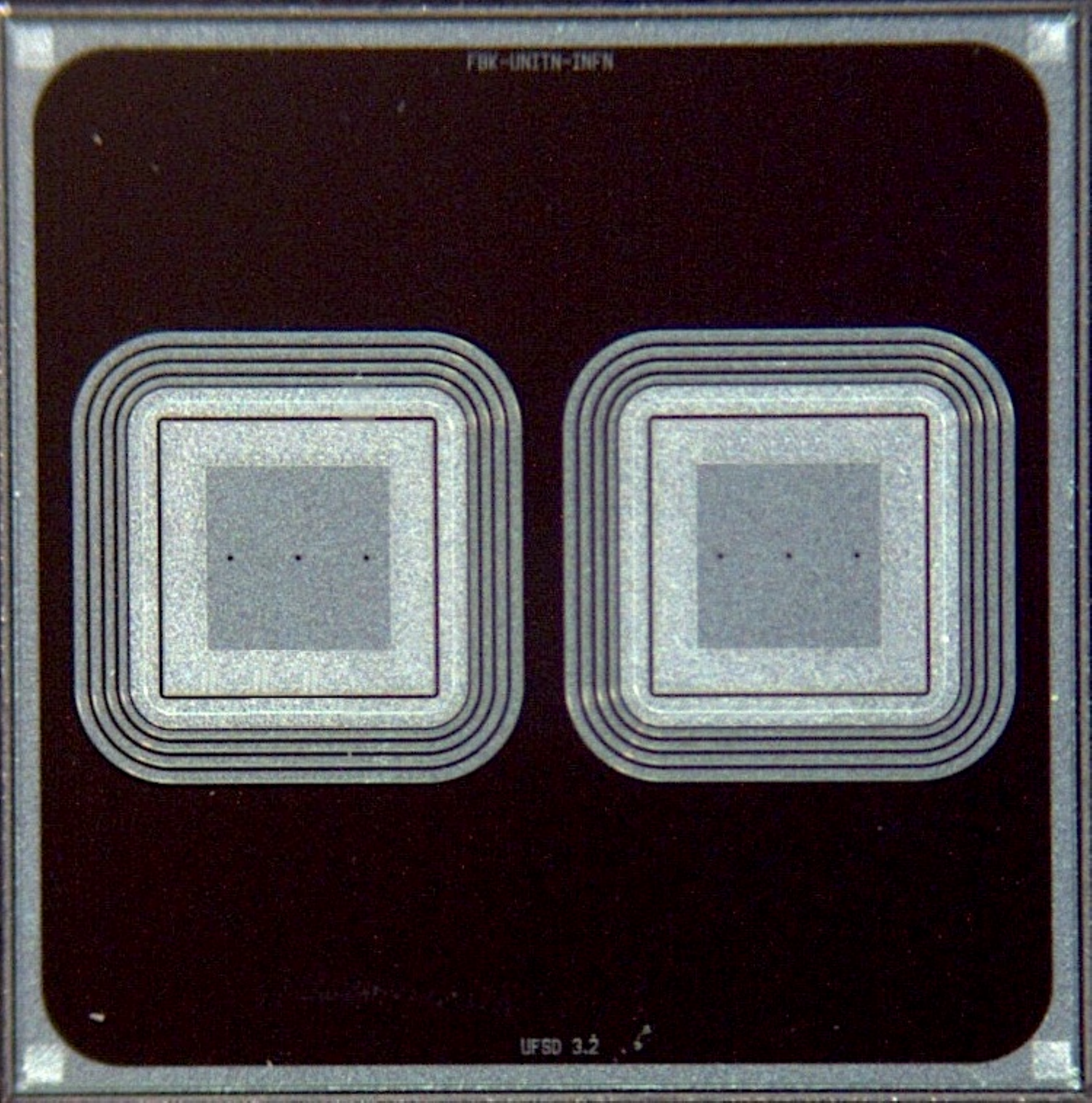}}\quad
        \centering%
        \subfigure[\label{fig:50L}]%
          {\raisebox{-0.0cm}{
          	\includegraphics[height=3.cm]{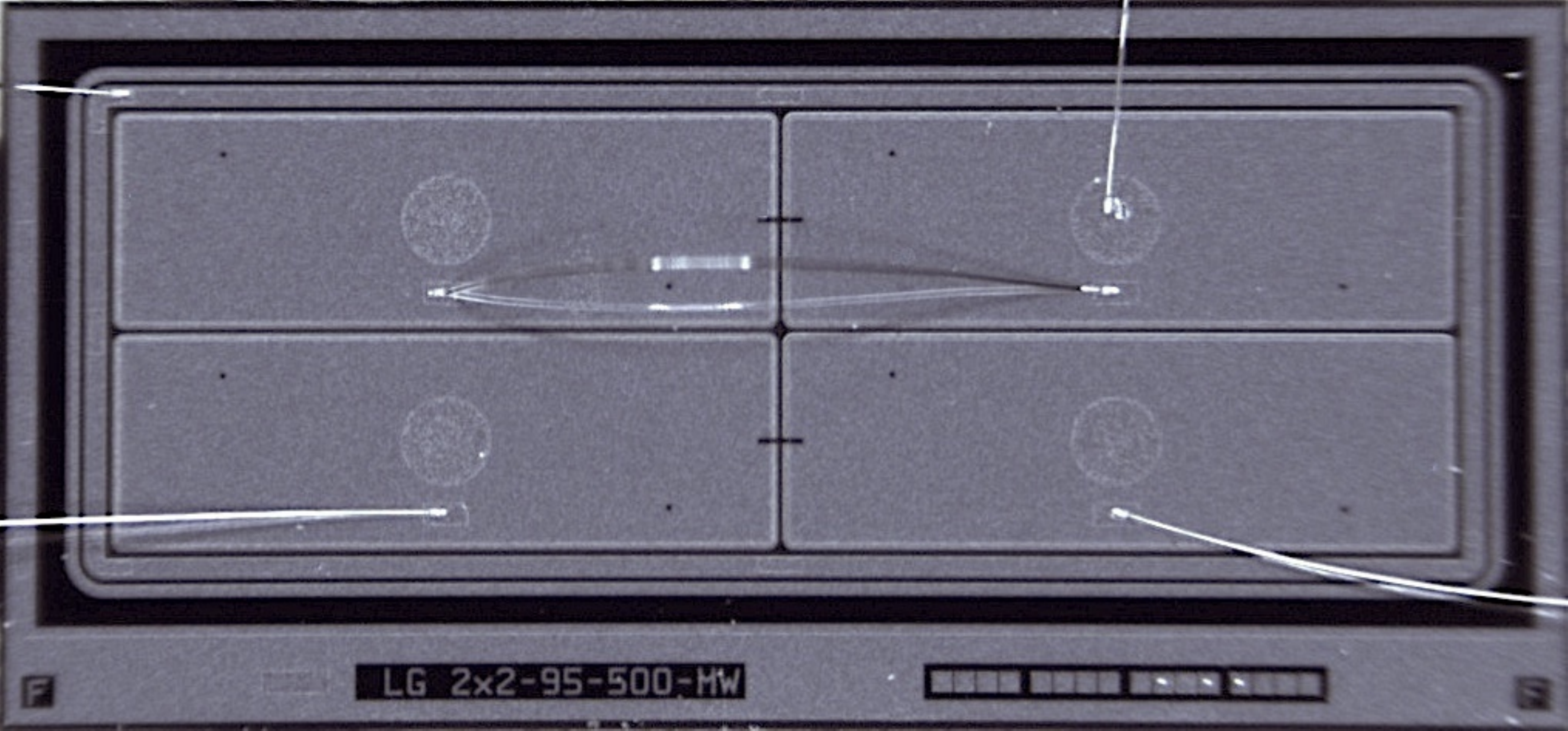}}}
          	\centering%
        \caption{(a) Picture of the FBK25 detector; on the left side the single 1 $\times$ 1 $\text{mm}^2$ UFSD tested and on the right an equivalent detector but without any gain layer, used for characterization measurements. (b) Picture of the HPK50 matrix with the four 1 $\times$ 3 $\text{mm}^2$ UFSD pads visible; only the bottom left has been readout in the measurements reported in this paper.}
        \label{fig:sensors}
\end{figure}

The tested UFSD came from two manufactures, Hamamatsu Photonics K.K. (HPK, Japan) and Fondazione Bruno Kessler (FBK, Italy). The one from HPK, hereinafter denoted HPK50, is a matrix of 4 pads (see Fig.\ref{fig:50L}), with a thickness of 50 $\mu$m; only one pad has been tested at the beam line. This device has been used as a comparison with previous measurements to validate the analysis and results.
Two UFSDs were instead manufactured from FBK, hereinafter denoted FBK25 and FBK35, with a nominal thickness of 25 $\mu$m and 35 $\mu$m, respectively; they represent the first thin wafers produced in 2020 by FBK\footnote{This production is called EXFLU0~\cite{exflu}. Before this production, the FBK UFSDs have never been thinner than 45 $\mu$m.}. In Figure \ref{fig:2535L}, a picture of FBK25 is shown (similar for FBK35).\\
\begin{table}[ht]
\begin{center}
\begin{minipage}{\textwidth}
\caption{Characteristics of the UFSDs under test.} \label{tab:lgad_char}
\begin{tabular*}{\textwidth}{@{\extracolsep{\fill}}lccccc@{\extracolsep{\fill}}}
\toprule
 & Area & Thickness & V$_{bd}$ & Voltage applied  & Gain\\
\midrule
FBK25  & 1 $\times$ 1 $\text{mm}^2$ & 25 $\mu$m & 127.3 $\pm$ 0.1 V & 75-120 V  & 13-57\\
FBK35 & 1 $\times$ 1 $\text{mm}^2$ & 35 $\mu$m & 260.7 $\pm$ 0.2 V & 165-240 V  & 10-49\\
HPK50 & 1 $\times$ 3 $\text{mm}^2$ & 50 $\mu$m & 253.0 $\pm$ 0.2 V & 200-245 V  & 26-61\\
\botrule
\end{tabular*}
\end{minipage}
\end{center}
\end{table}
All the detectors tested have been previously completely characterized at the INFN Bologna laboratories. 
The main characteristics of the sensors are reported in Table \ref{tab:lgad_char}. The breakdown voltage (V$_{bd}$) has been extracted from I-V measurements using the methods of LD and ILD (Logarithmic Derivative and Inverse LD, \cite{2017Klanner}).
The internal gain has been extracted considering the charge released from a MIP on the Detector Under Test (DUT) at a certain voltage and comparing it with the charge released in an equivalent detector but without any internal gain.\\

 \begin{figure}
        \centering%
          {\includegraphics [width=6cm]{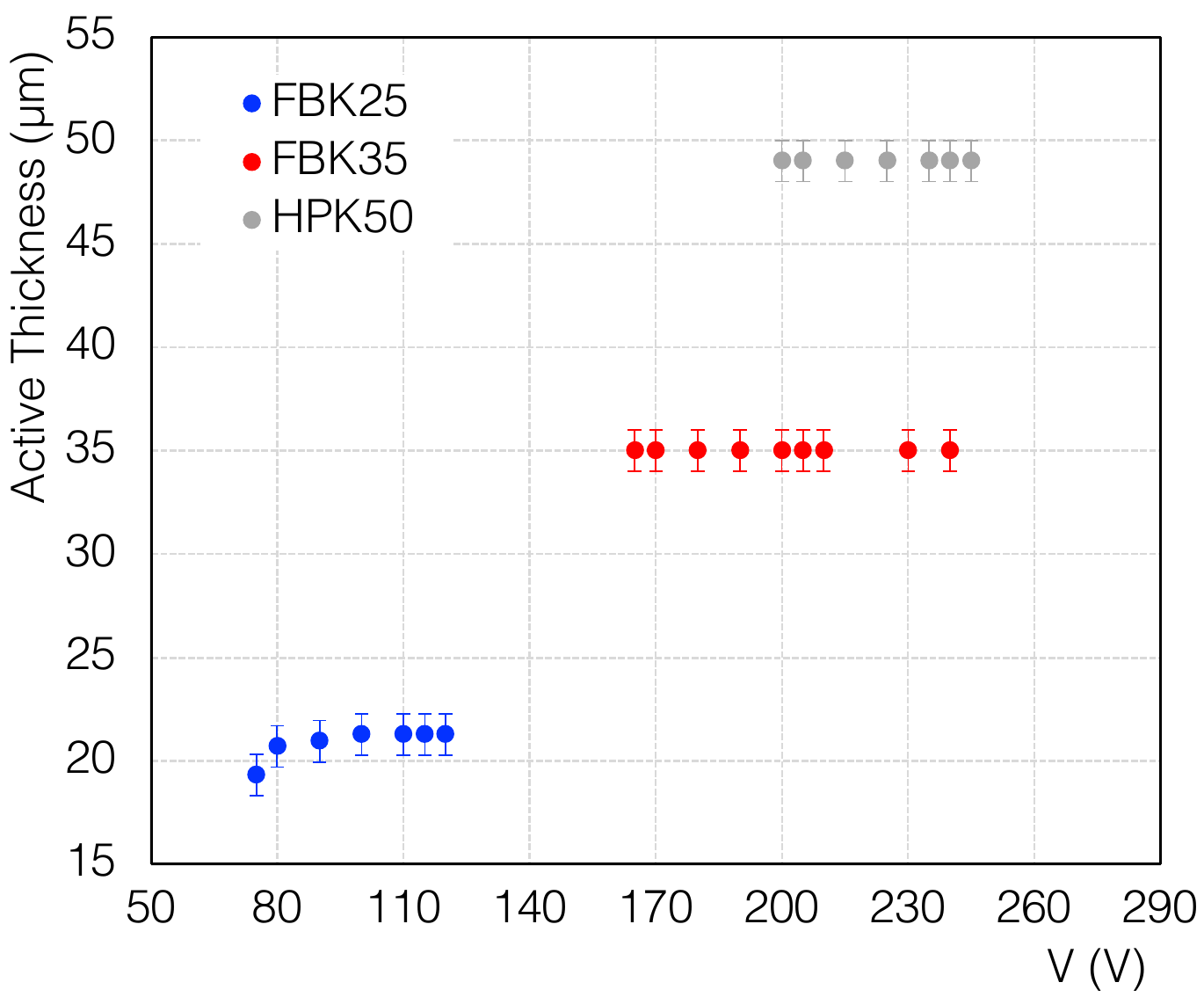}}
        \caption{Active thicknesses extracted as a function of the applied voltages.}
          \label{fig:thickV}
\end{figure}
The actual thickness of the DUTs has been evaluated by performing a Capacitance-Voltage measurement\footnote{Simply extracting the thickness from the capacitance measurement as a function of the bias voltage, with the approximation of a planar detector geometry.}. In Figure \ref{fig:thickV} the extracted active thicknesses are reported. As can be noticed, while the HPK50 and FBK35 are independent from the voltage in that range, the FBK25 is instead still depleting and it reaches a plateau of 21 $\mu$m only at around 100 V.
Notice that the final active thickness is slightly lower than what expected.
The different outcome of the FBK25 is due to the low resistivity of this sensor, connected to the very high doping of the bulk, resulting also in a non-uniform electric field of the drift region.

\subsection{Beam test setup and electronics}
\label{sec:set_el}
The time resolution of the UFSDs has been studied at the T10 beamline at PS-CERN in November 2021. The beam was mainly composed of protons and pions with a momentum of +12 GeV/c.
The trigger for the data acquisition was given by the coincidence of the self-trigger from the UFSDs.
For each data acquisition up to 4 carrier boards were aligned to the beam in a telescope frame and the whole setup was enclosed in a dark environment box at room temperature.

The UFSDs have been mounted on different front-end boards.
The board used for the HPK50 was followed by a C2 broadband (2 GHz) and low-noise current amplifier\footnote{\href{https://cividec.at/electronics-C2.html}{Cividec datasheet}} with a gain of 190.
For both the FBK sensors instead the board used, V1.4-SCIPP-08/18, contained a wide bandwidth (2 GHz) and low noise inverting amplifier with a measured amplification of factor 6. The board was followed by a second amplification stage, with a gain factor of around 13 and 14 respectively for the FBK25 and FBK35 \footnote{The second amplifier used for the FBK25 and FBK35 were the minicircuit LEE39+ (\href{https://www.minicircuits.com/pdfs/LEE-39+.pdf}{LEE39+ datasheet}) and Gali52+ (\href{https://www.minicircuits.com/pdfs/GALI-52+.pdf}{GALI52+ datasheet}) respectively.}.

Up to 4 amplified signals were sent to a  4 GHz LeCroy WaveRunner 9404M-MS oscilloscope\footnote{\href{https://teledynelecroy.com/oscilloscope/waverunner-9000-oscilloscopes/waverunner-9404m-ms
}{Lecroy WaveRunner datasheet}}, with 20Gs/s samplig rate and 4 GHz-8 bit vertical resolution. The contribution of the oscilloscope time resolution to the measured one was negligible.

 \begin{figure}[h!!]
        \centering%
          {\includegraphics [width=8cm]{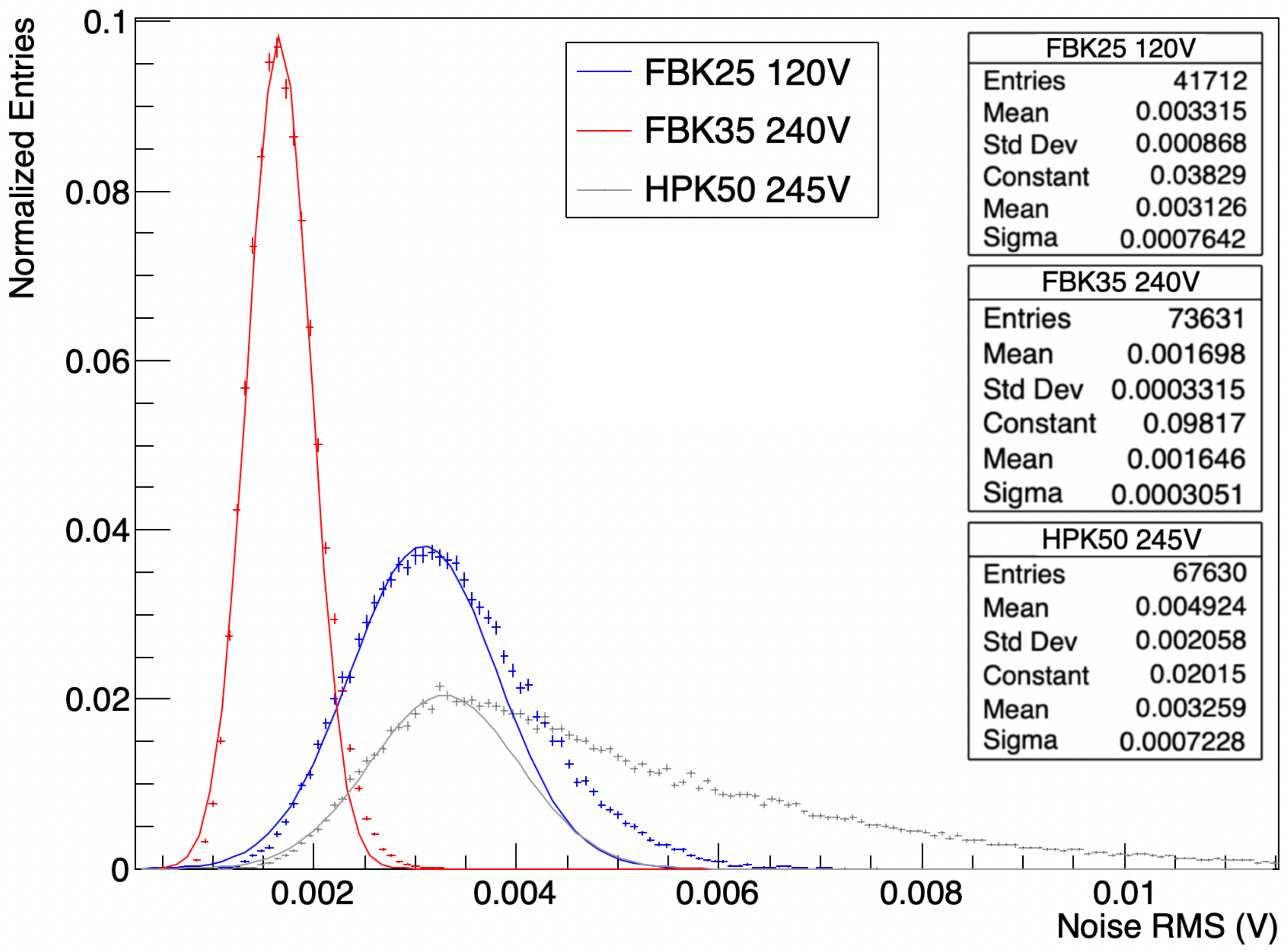}}
        \caption{RMS noise spectra for the three different LGADs tested. The distributions have been fit with a gaussian, see the text for details.}
          \label{fig:rms}
\end{figure}
To evaluate the noise of each DUT, its RMS has been evaluated; in particular for the analysis offline a time window before the signals has been considered in order to have a direct measurement of the noise in the beam test setup. In Figure \ref{fig:rms} the RMS spectra are reported. As can be seen, the FBK35 has a lower noise with respect to the FBK25.
For the FBK25 and FBK35 and all the voltages reported, the distributions have a Gaussian shape, with only a moderate non-gaussian right tail that instead is much more pronounced for the HPK50. 
The distribution mean of the three DUTs is quite stable throughout all the data taking, ranging between 1 mV and 4 mV.

\section{Results} 
\label{sec:results}
The data analysis was performed following similar procedures to those reported in \cite{2019Carne}. In particular, thanks to the oscilloscope readout, the completed signal waveforms were recorded and analyzed. It was then possible to use the Constant Fraction Discriminator (CFD) method to extract the DUTs time resolutions.
Moreover, to filter-out the high-frequency noise, a smoothing of the LGAD signal was applied with a moving average of four consecutive points.

\begin{figure}[h!!]
        \centering%
          {\includegraphics [width=8cm]{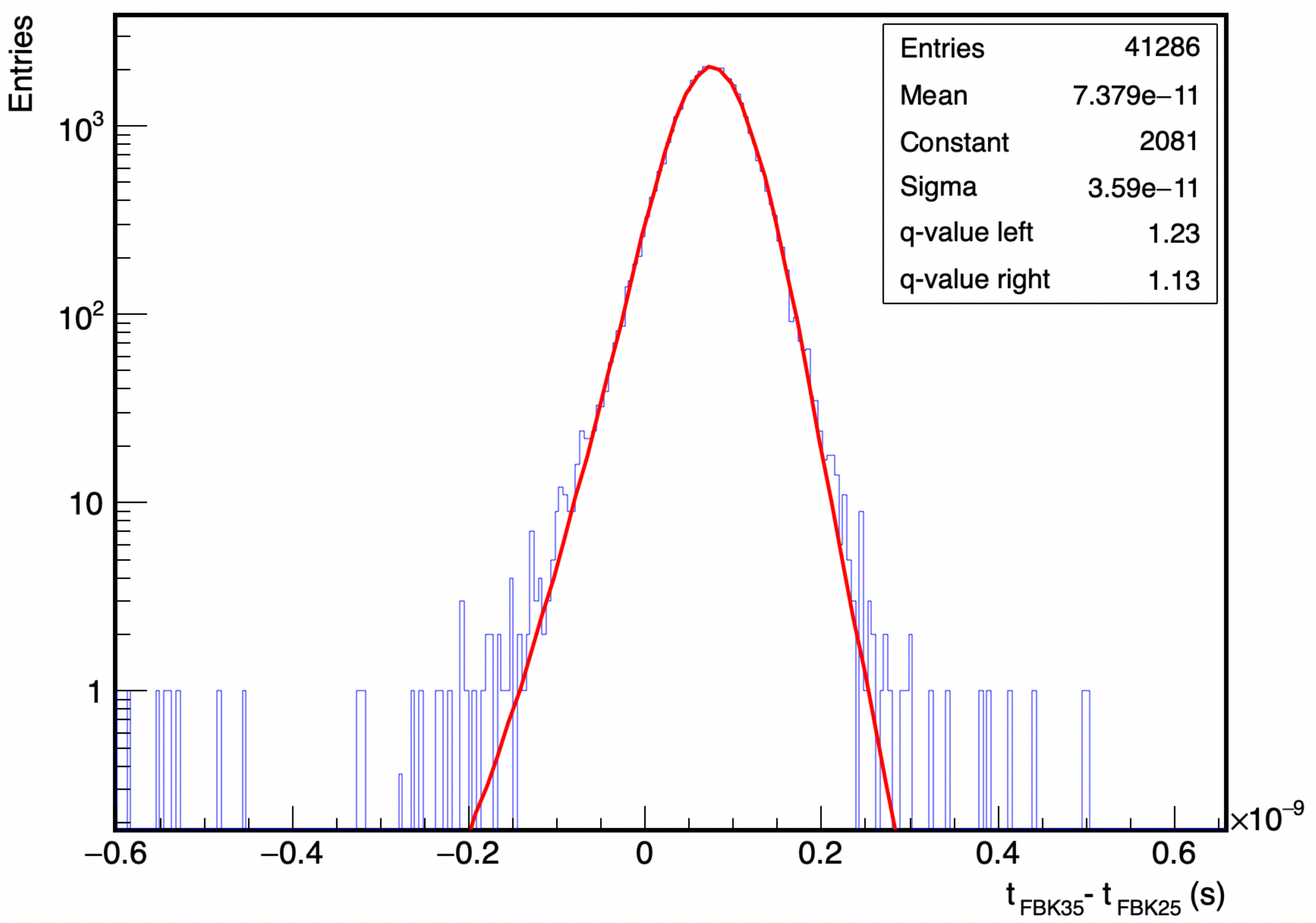}}
        \caption{Time difference distribution between FBK35 and FBK25 samples, operated respectively at 240 V and 120 V, and with a CFD of 70$\%$. The extracted time resolutions is in this case around 25 ps for both the UFSD sensors. The distribution has been fitted with a q-Gaussian function. In this example, the tails account for the 2.5\% of the measures.}
          \label{fig:Q_Gaussian}
\end{figure}

\begin{figure}[htbp]
        \centering%
        \subfigure[\label{fig:25um_res_cfd}]%
          {\includegraphics [height=4.5cm] {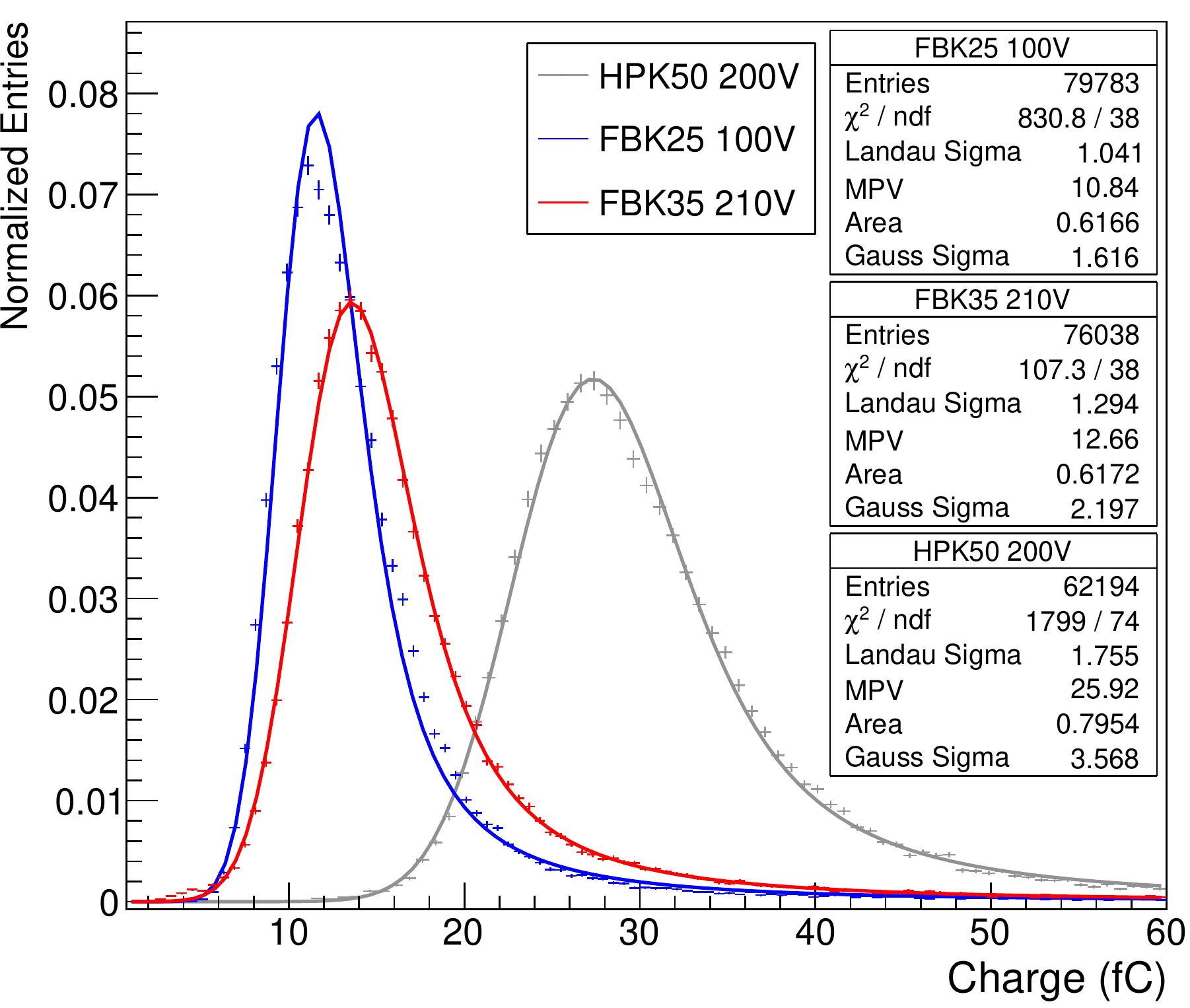}}\quad
        \centering%
        \subfigure[\label{fig:35um_res_cfd}]%
          {\raisebox{-0.0cm}{
          	\includegraphics[height=4.5cm]{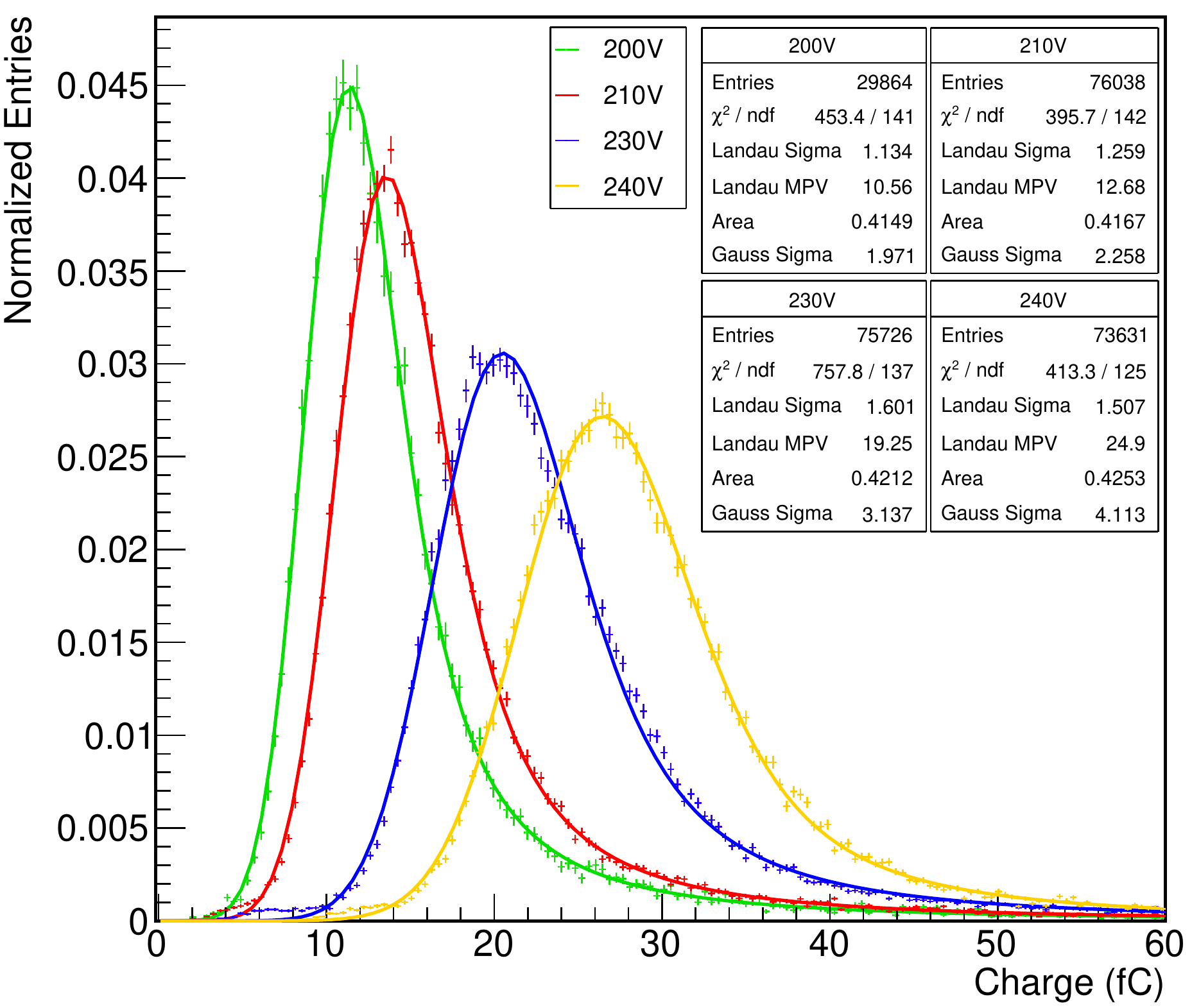}}}
          	\centering%
        \caption{(a) Charge distributions obtained for the three thicknesses at a similar gain ( $\sim$23). (b) Charge distribution  for the
FBK35 as a function of different applied voltages. The distributions are fitted with a convolution of a Gaussian and a Landau functions.}
        \label{fig:landau}
\end{figure}

To extract the time resolution of a single UFSD, a system with  three sensors and three differences between the arrival time of each pair of detectors has been considered. In Figure \ref{fig:Q_Gaussian} an example of the time difference between FBK35 and FBK25 is reported. The distribution has been fitted with an asymmetric q-Gaussian function to take into account that the arrival time distribution for a single sensor has a Gaussian shape with a small tail towards late times. The sigma extracted from the fit has then been used to obtain the final time resolution of the three UFSDs at a given voltage and CFD.\\
In Figure \ref{fig:landau} the measured charge distributions are shown. The ratio between the width of the distributions and the Most Probable Values(MPVs) decreases as a function of the thickness. As expected, the amplitude increases both as a function of the thickness and voltage; moreover, the fluctuations in the number of electrons-holes created in the silicon detector decrease with larger thickness, in accordance with \cite{2011Meroli} and \cite{2017Riegler}.
 \begin{figure}[htbp]
        \centering%
        \subfigure[\label{fig:25um_res_cfd}]%
          {\includegraphics [height=4.5cm] {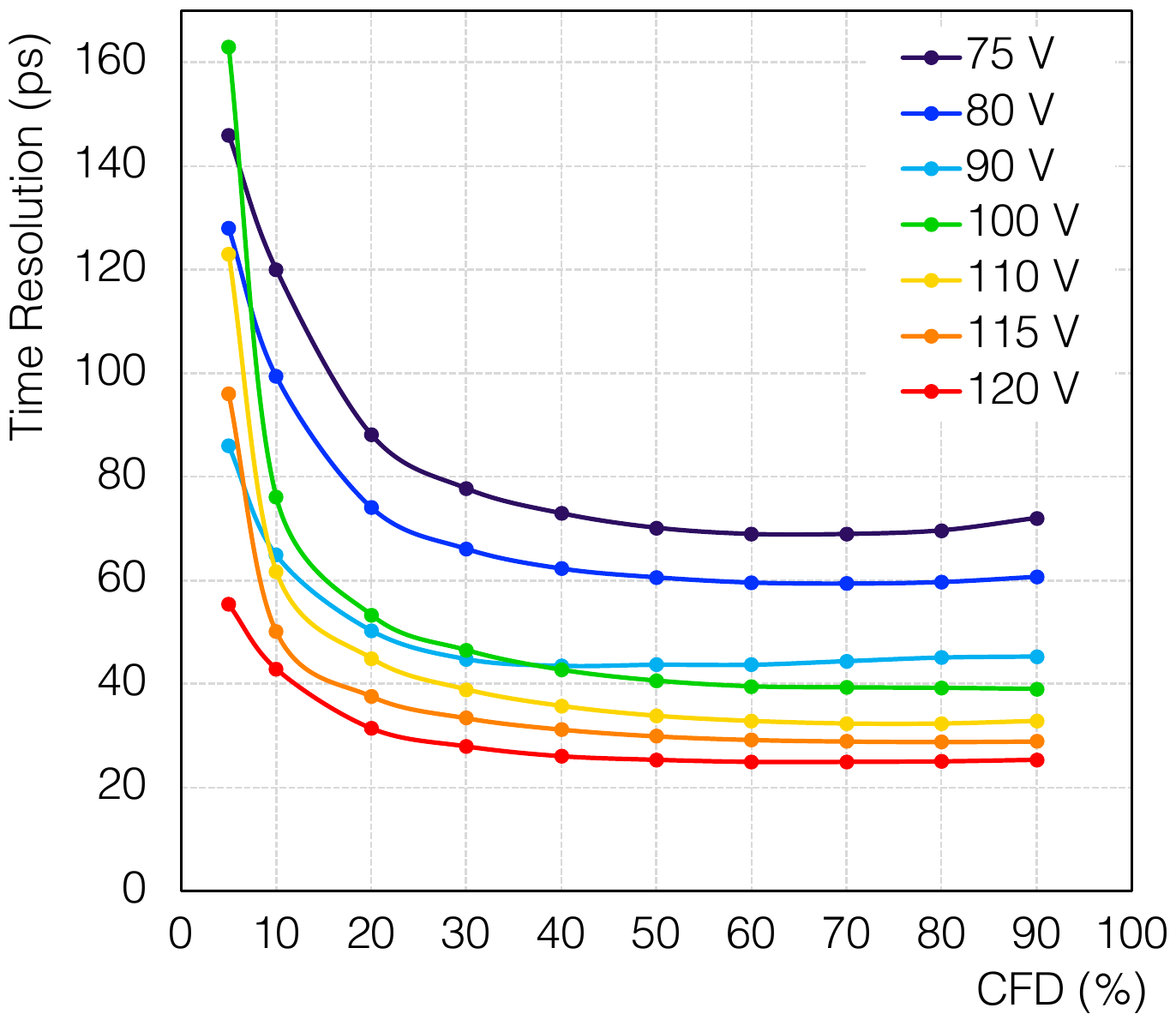}}\quad
        \centering%
        \subfigure[\label{fig:35um_res_cfd}]%
          {\raisebox{-0.0cm}{
          	\includegraphics[height=4.5cm]{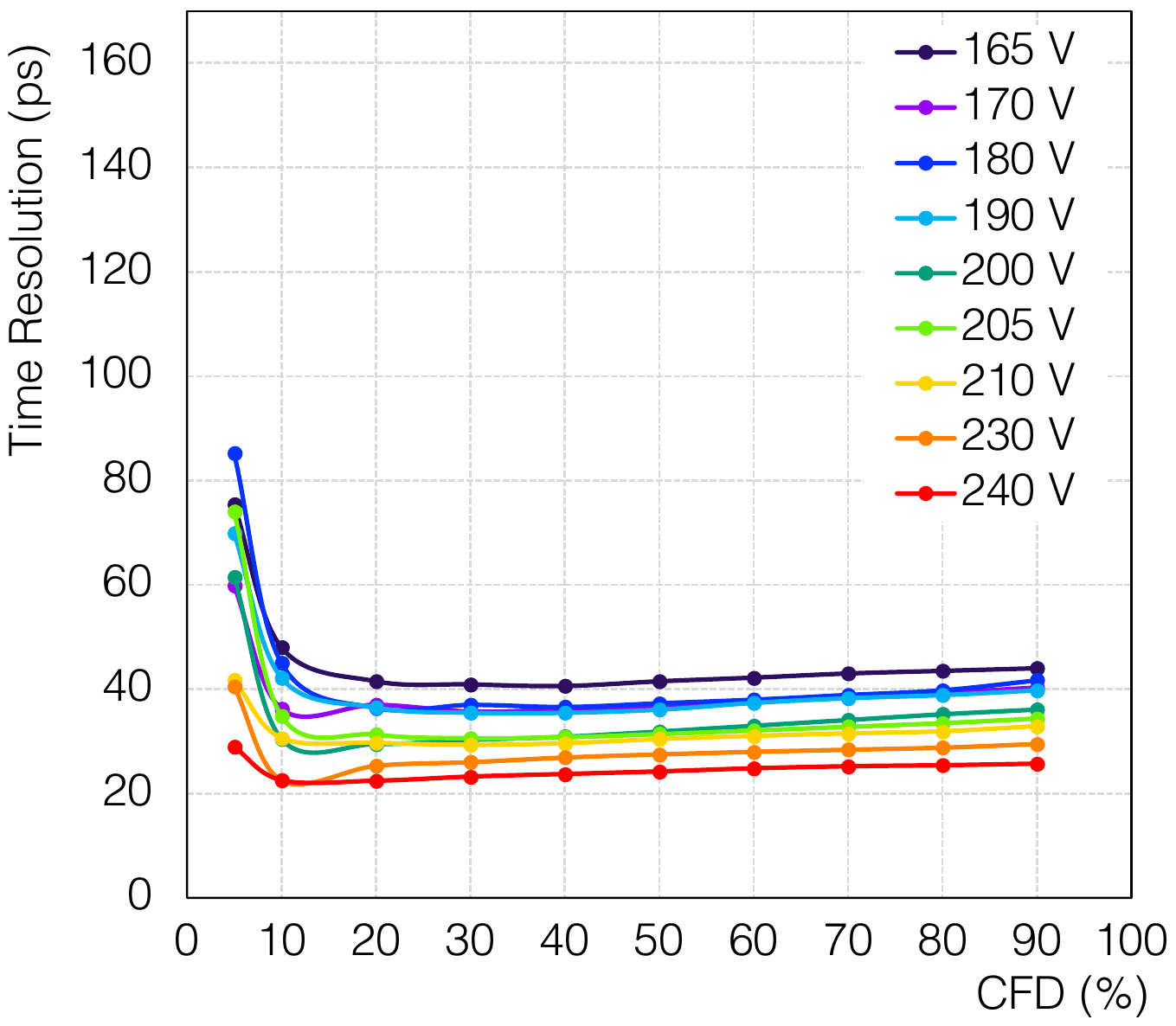}}}
          	\centering%
        \caption{Measured time resolution versus the CFD threshold for several applied voltages for (a) FBK25 and (b) FBK35. The errors have been estimated to be around 10$\%$  of the measured value for CFD between 20$\%$ and 80$\%$ (not shown in the plots for better visibility).}
        \label{fig:res_cfd}
\end{figure}
Figure \ref{fig:res_cfd} reports the measured time resolutions for each sensor  as a function of the CFD threshold for all the applied voltages.
For the HPK50 the trend and values are totally in agreement with the ones reported in \cite{2017Cartiglia, 2019Carne} for a 50 $\mu$m thick UFSD.
As expected, the time resolution improves with higher voltages, and the trend as a function of the CFD reflects the higher jitter contribution for very low CFD (smaller slew rates) \cite{2017Sadrozinski}. At higher CFD instead the time resolution is dominated by the Landau term, which is due to the non-uniform creation of electron–hole pairs along the particle path and which is expected to be smaller for thinner detectors. The trend of the measured values vs CFD  flattens for values larger than around 30$\%$ and 20$\%$, respectively for the FBK25 and the FBK35 sensors. 
To understand the trends reported, different contributions have to be taken into account. In particular the two main ones: the jitter and Landau terms. At low CFD values, the jitter contribution is dominant while for higher CFD values the landau terms start to be the dominant one. In turn they depend on the thickness, S/N and slew rate; these contributions are affected both by electronics and detector, their coupling and their shielding against outer noise\cite{2017Sadrozinski}. 
The S/N ratio has then been evaluated for the DUTs; a value between 6-24 and 16-63 (depending on the different applied voltage) for the FBK25 and FBK35 sensors have been found respectively.
The jitter has been evaluated for each DUT and the intrinsic time resolution has been extracted: $\sigma_\text{intrinsic}^2=\sigma_\text{measured}^2-\sigma_\text{jitter}^2$.\\

\begin{figure}
        \centering%
          \subfigure[\label{fig:res_ch}]%
          {\includegraphics [width=5cm] {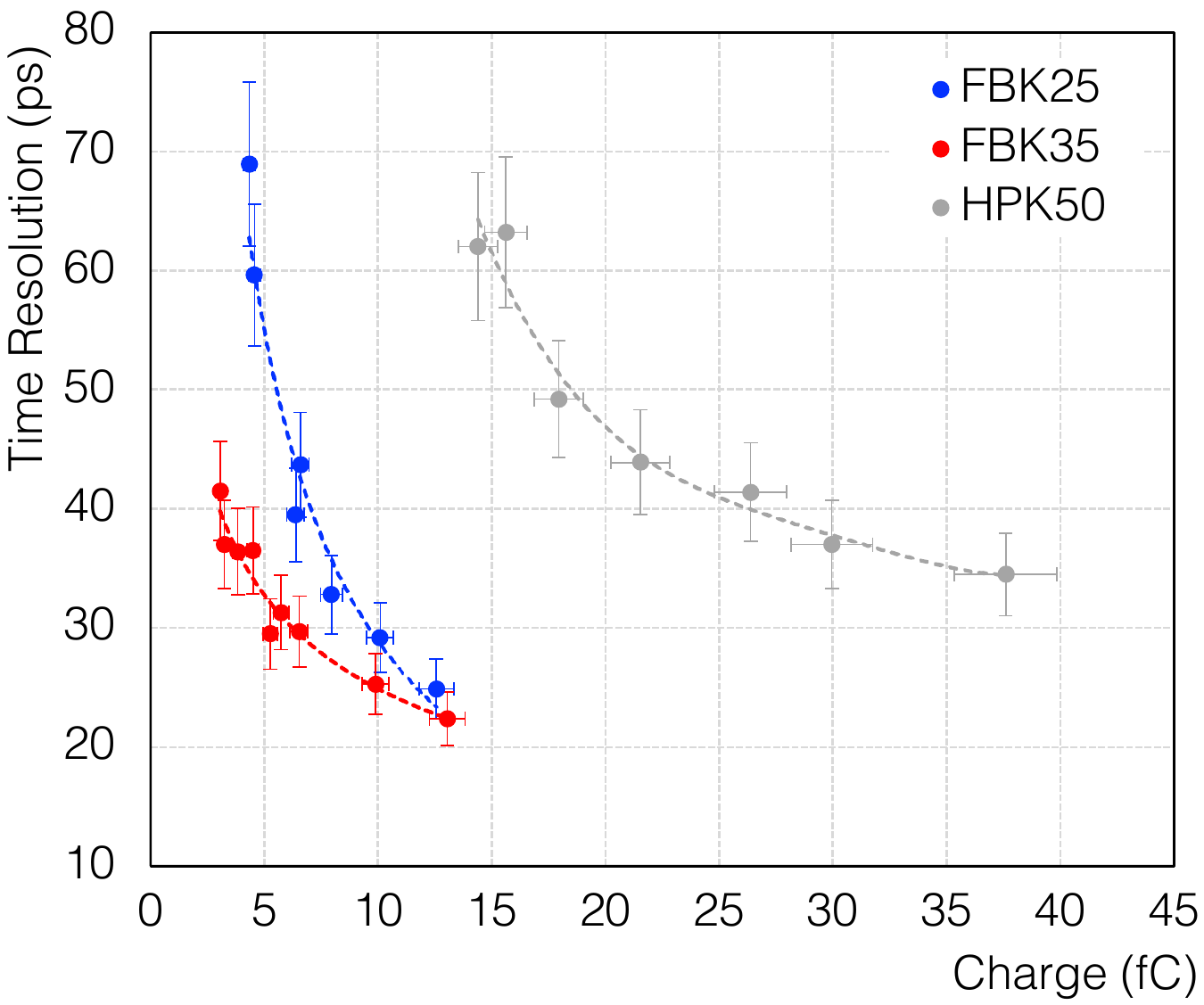}}\quad
        \centering%
        \subfigure[\label{fig:res_E}]%
          {\raisebox{-0.0cm}{
          	\includegraphics[width=5cm]{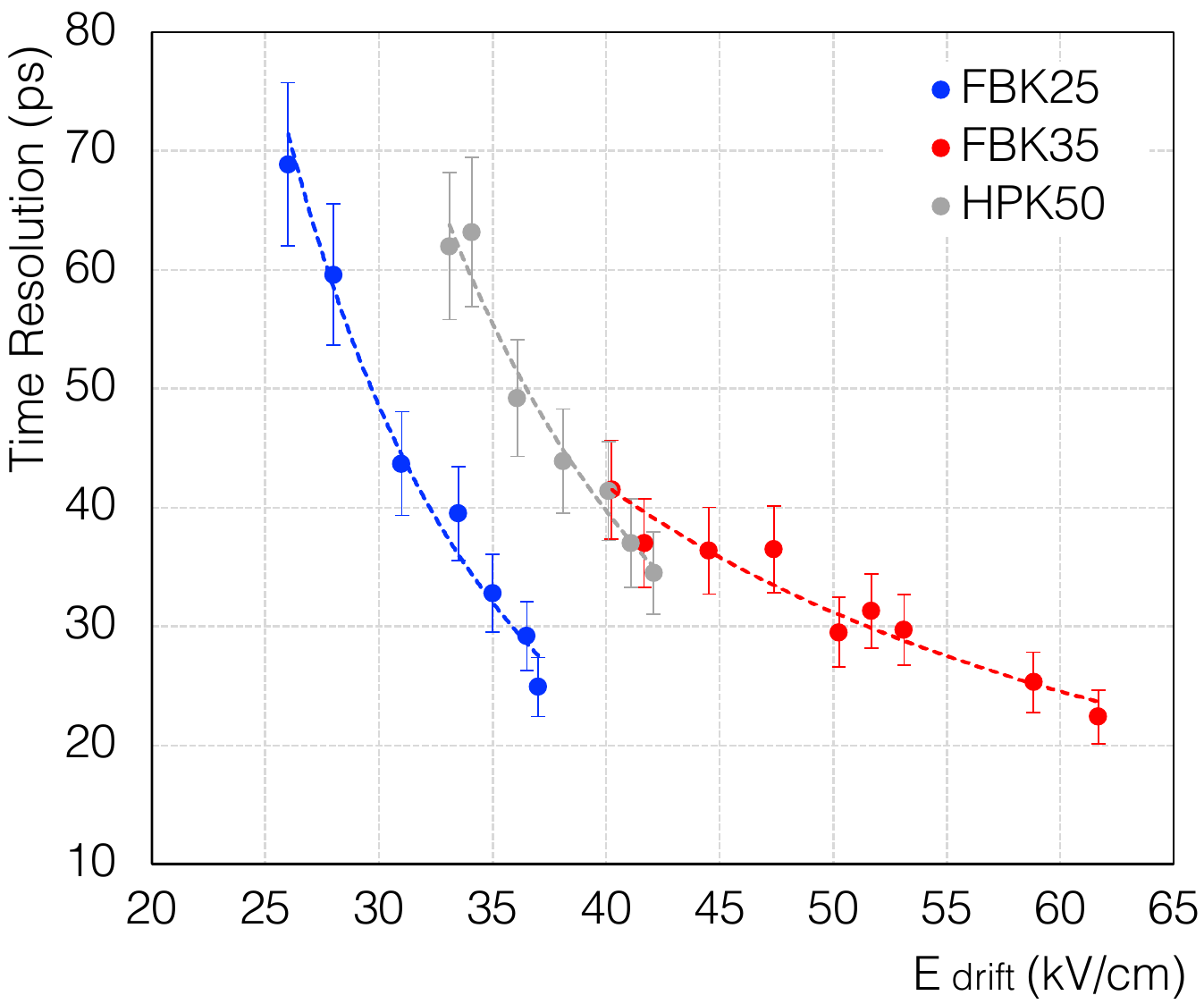}}}
          	\centering%
        \caption{\label{fig:res_chE} Measured time resolution results from the beam test as a function of the (a) charge and (b) drift electric field (E$_\text{drift}$) for all the UFSDs: FBK25, FBK35 and HPK50 for a CFD of 60\%, 20\% and 50\%, respectively. The errors for the measured time resolution have been estimated as 10$\%$ of the value.  The lines are included to guide the eye.}
\end{figure}

In Figure \ref{fig:res_chE} the measured time resolutions for a fixed CFD (the one that minimizes the time resolution) are reported as a function of (a) charge and (b) drift electric field\footnote{N.B. With drift electric field we refer to the electric field inside the silicon bulk (drift region), not to the electric field in the gain region. The drift electric field has been calculated for the 35 and 50 $\mu$m thick sensors as the difference between the applied voltage and the voltage used to deplete the gain layer, divided for the LGAD thickness. Instead for the FBK25, because of the non uniformity of the electric field due to the high doping, a mean value has been extracted using a Weightfield2(WF2)\cite{CENNA2015149} simulation.} for the three DUTs. The time resolution improves with larger charge (as expected due to larger gain) and drift electric field (as expected due to both larger gain and higher velocity of carriers in the drift region).

\begin{figure}
        \centering%
          \subfigure[\label{fig:res_gain}]%
          {\includegraphics [width=5cm] {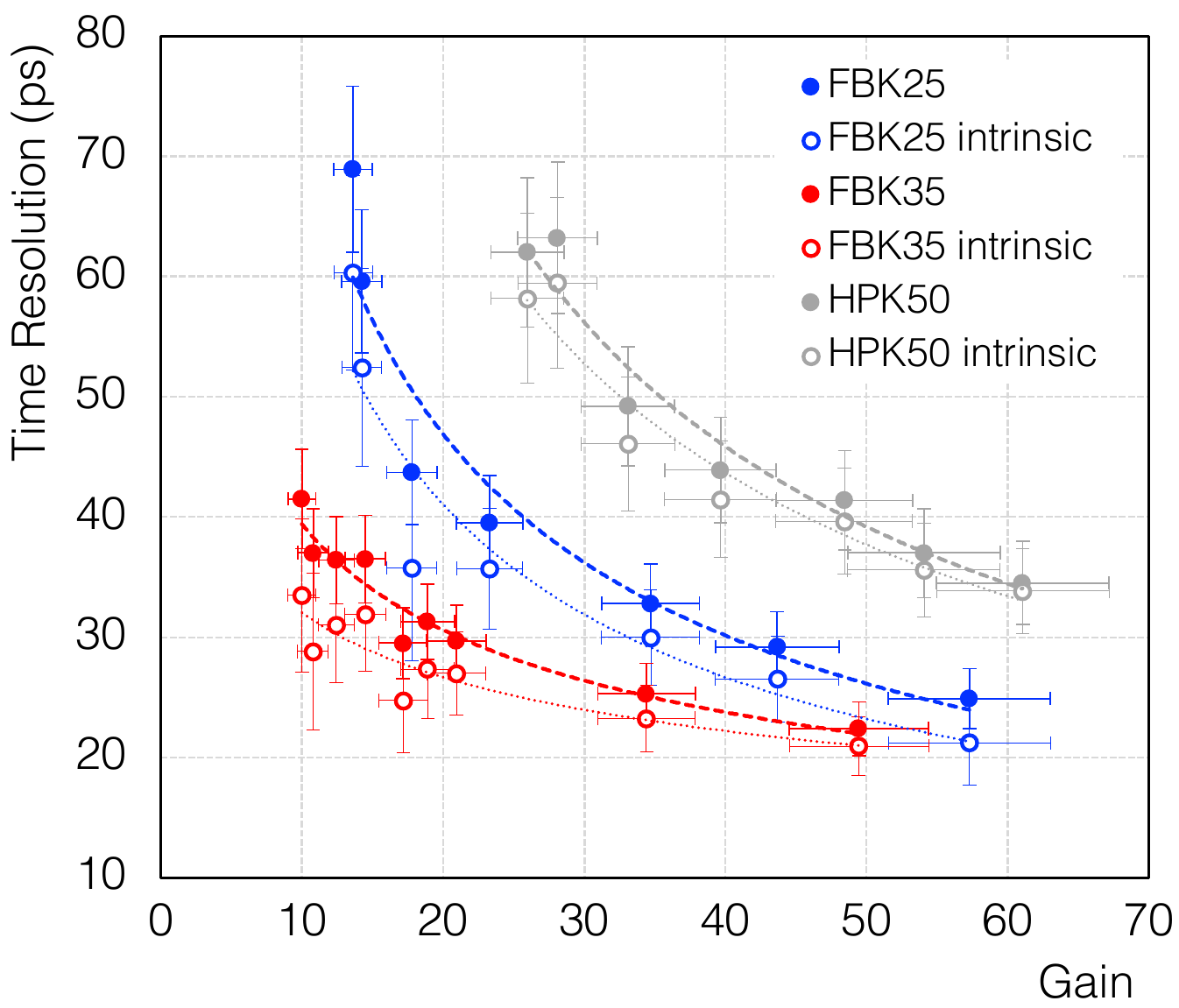}}\quad
        \centering%
        \subfigure[\label{fig:res_V}]%
          {\raisebox{-0.0cm}{
          	\includegraphics[width=5cm]{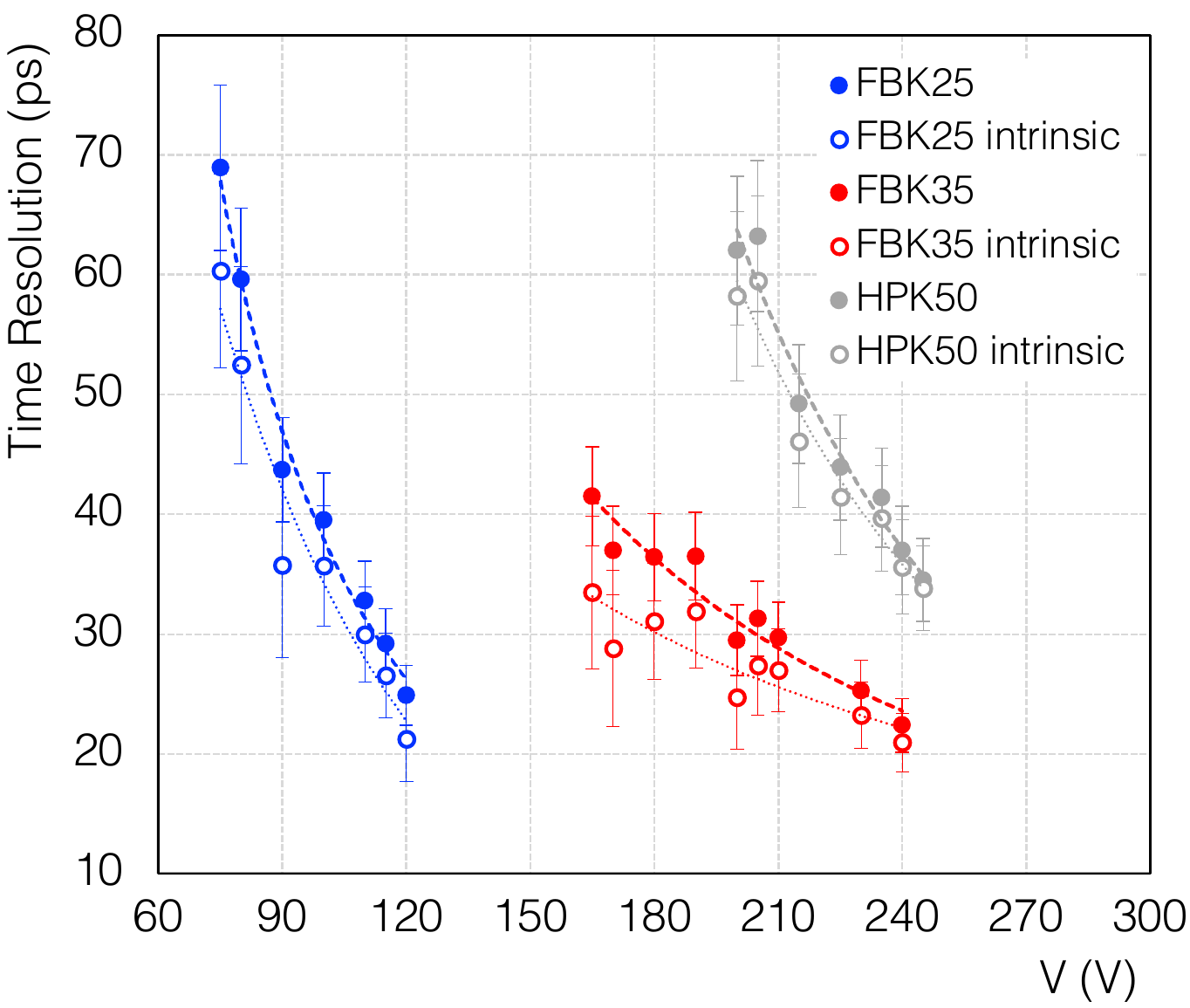}}}
          	\centering%
        \caption{\label{fig:res}Measured (full circle) and intrinsic (empty circle) time resolution from the beam test as a function of the (a) gain and (b) applied voltage for all the UFSDs: FBK25, FBK35 and HPK50 for a CFD of 60\%, 20\% and 50\%, respectively. The errors for the measured time resolution are estimated to be around 10$\%$ of the value; the error on the gain are estimated to be around the 10$\%$ of the value. The lines are to guide the eye.}
\end{figure}

In Figure \ref{fig:res} the measured and intrinsic time resolutions for a fixed CFD are reported as a function of gain (a) and voltage (b). 
In the figure, the three sensors with different investigated thicknesses are compared. As expected, the time resolution improves for higher gain (voltages) for all the detectors.

A time resolution of $\sim$34 ps has been measured for the HPK50, which confirms previous results \cite{2017Cartiglia, 2019Carne}.
As expected, a better time resolution has been reached with the two thinner UFSDs due to a smaller Landau term. 
However, the measurements of the two samples FBK25 and FBK35, although compatible within the uncertainties, have a trend towards better results for FBK35 w.r.t. FBK25.
The behaviour of the FBK25 can be due to several reasons, including a worse S/N ratio (due also to the higher noise RMS, see Figure \ref{fig:rms}) and a bulk doping not optimized for timing measurements. Future production of thin sensors with almost intrinsic doping of the substrate are foreseen.
On the other side, the time resolution for the FBK35 showed results in agreement with the expectation; the time resolution reached can be compared with the one obtained in \cite{2021Jadhav} of 25.6 ps for the same thickness.

\begin{figure}
        \centering%
          {\includegraphics [width=9cm]{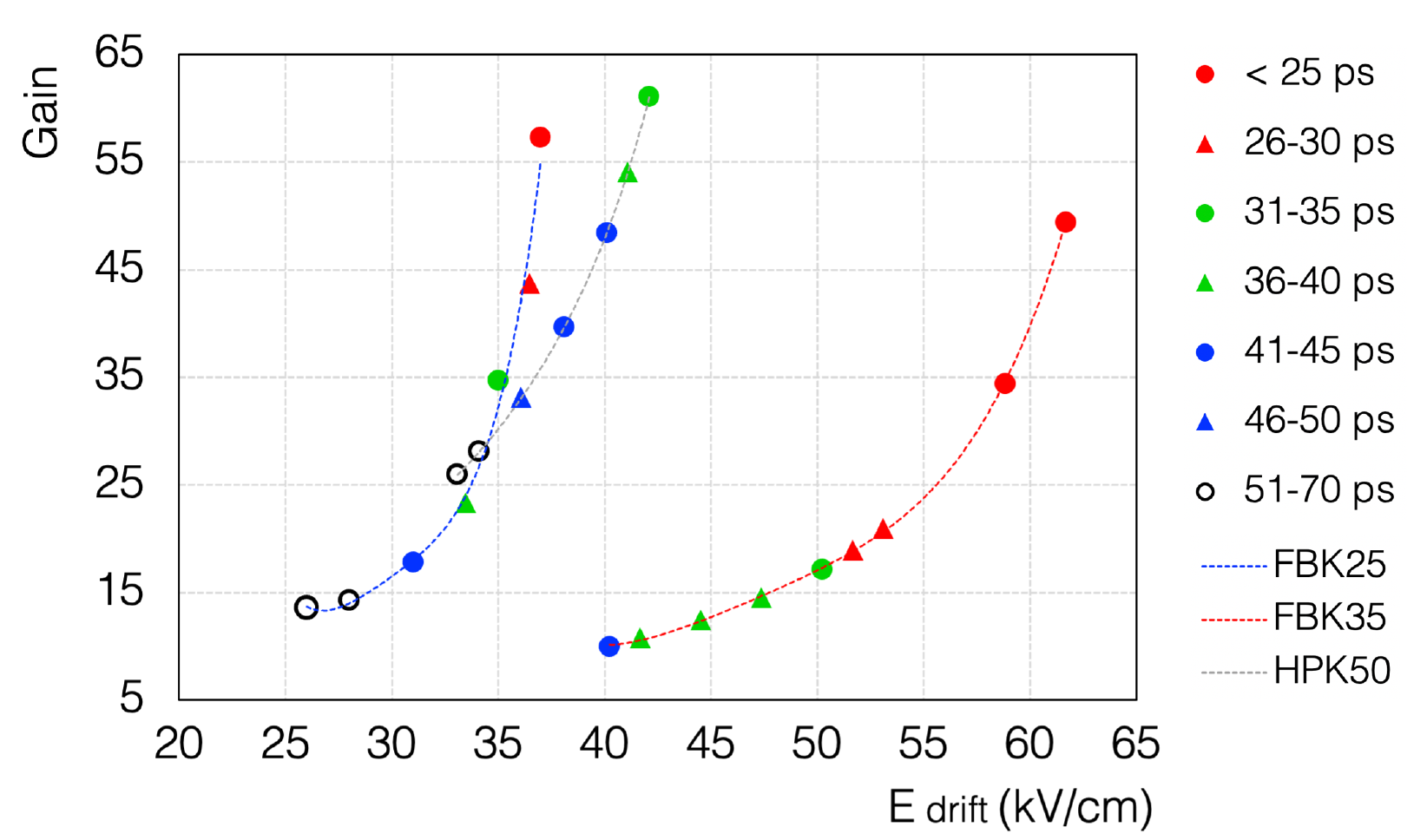}}
        \caption{Measured time resolution for the chosen CFD as a function of gain and drift electric field (E$_\text{drift}$) for FBK25, FBK35 and HPK50. Each line joins the measurements of a given sensor; the colors indicate a give time resolution interval.}
          \label{fig:timeElGain}
\end{figure}

A summary plot of time resolution as a function of both gain and drift electric field is reported in Figure \ref{fig:timeElGain} for the UFSDs tested. As expected \cite{2020_Roberta}, it indicates that to obtain a good time resolution a combination of both gain and drift electric field is necessary.

\begin{table}[ht]
\begin{center}
\begin{minipage}{\textwidth}
\caption{Time resolution for FBK25, FBK35 and HPK50 for a given voltage (or gain) obtained in a beam test setup at room temperature.} \label{tab:res}
\begin{tabular*}{\textwidth}{@{\extracolsep{\fill}}lccc@{\extracolsep{\fill}}}
\toprule
 & Voltage applied & Gain & Time resolution\\
\midrule
FBK25 & 120 V & 57 $\pm$ 4 & (25 $\pm$ 3) ps\\
FBK35 & 240 V & 49 $\pm$ 5 & (22 $\pm$ 2) ps\\
HPK50 & 245 V & 61 $\pm$ 3 & (34 $\pm$ 3) ps\\
\botrule
\end{tabular*}
\end{minipage}
\end{center}
\end{table}

In Table \ref{tab:res} the best time resolutions reached for the three detectors tested are summarized.

 \begin{figure}
        \centering%
          \subfigure[\label{fig:sim_res_gain}]%
          {\includegraphics [width=5cm] {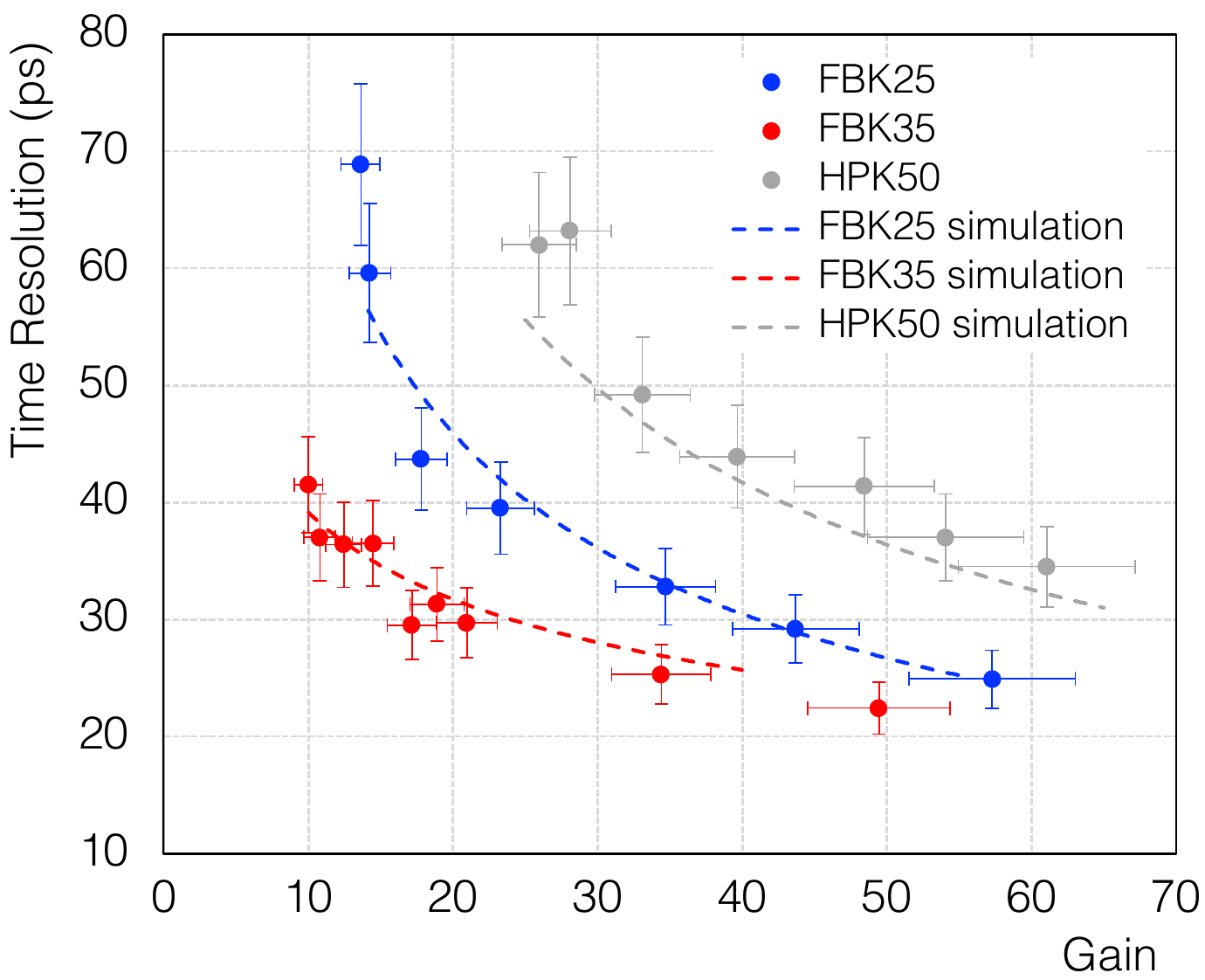}}\quad
        \centering%
        \subfigure[\label{fig:sim_res_V}]%
          {\raisebox{-0.0cm}{
          	\includegraphics[width=5cm]{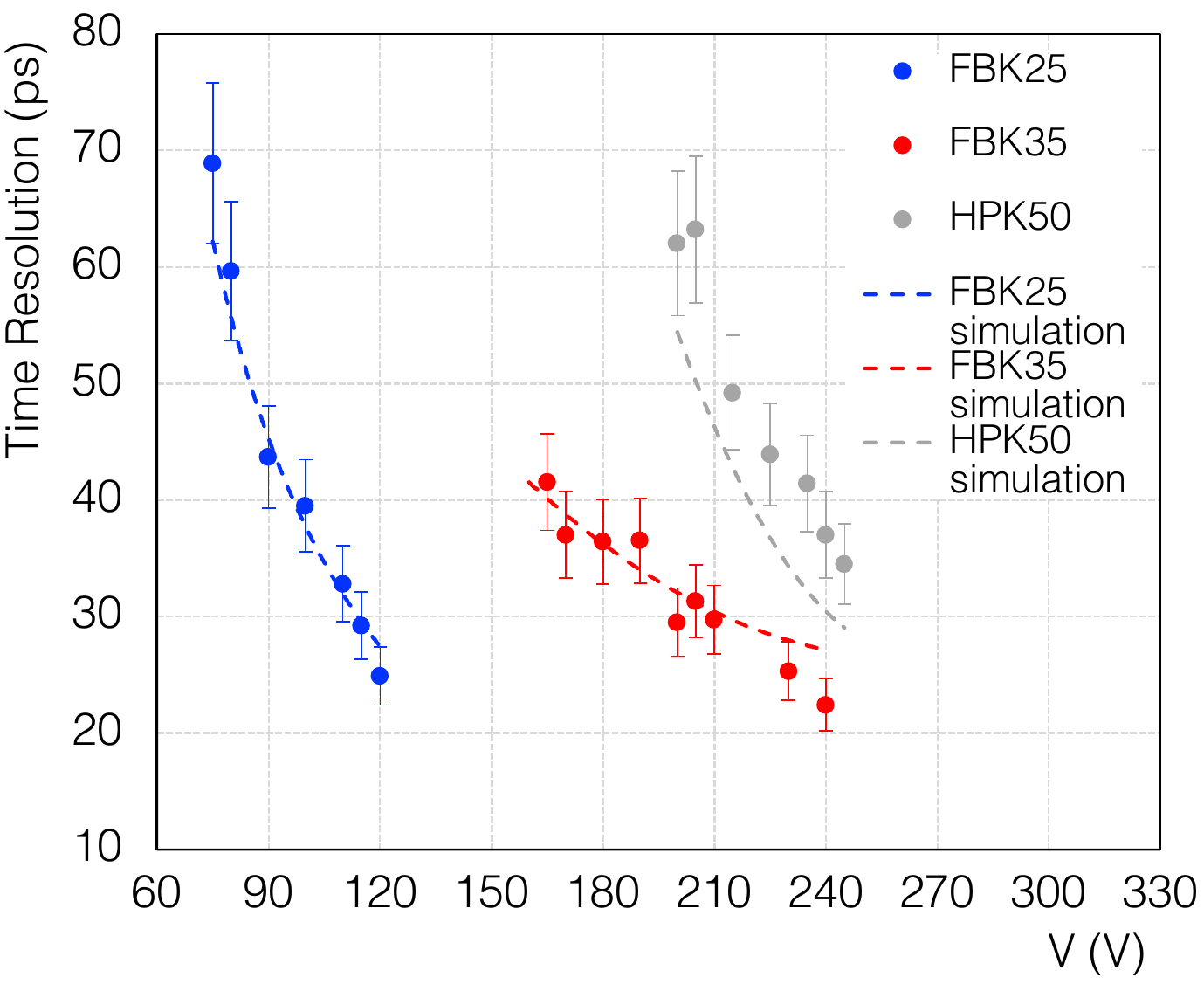}}}
          	\centering%
        \caption{\label{fig:Simulation}Measured (circle) and simulated (dotted lines) time resolution results from the beam test  and simulation with WF2, as a function of the (a) gain and (b) applied voltage for all the UFSDs: FBK25, FBK35 and HPK50 for a CFD of 60\%, 20\% and 50\%, respectively. The errors for the measured time resolution have been estimated as 10$\%$ of the value; the error on the gain has been estimated as the 10$\%$ of the value.}
\end{figure}

Finally the results have been also compared with simulation.
In Figure \ref{fig:Simulation} the measured time resolution obtained for all the LGADs is reported as a function of (a) gain and (b) applied voltage and compared with the one obtained from simulation, using WeightField2. The timing capabilities were evaluated applying the CFD method, using the same threshold adopted for the measured data. The time resolution of the different LGAD has been calculated by choosing the parameters of the simulation in order to reproduce the signal waveforms obtained at the beam test by the sensors connected to the electronics. This procedure has been tuned for one bias point for each of the 3 sensors and then extended to the others. In this way it is possible to compare the trend of data with what expected from the simulation with WF2.
As shown in the figure, the simulation reproduces accurately the evolution of the time resolution with gain (left pane) and bias voltage (right pane).

\section{Conclusions}\label{sec:conclusions}

In this paper, several measurements of time resolution for a new generation of UFSDs are reported. The results have been obtained  using a 12 GeV/c beam at CERN PS. An HPK UFSD of a standard thickness of 50 $\mu$m has been tested and used as reference; it reaches a time resolution of $\sim$34 ps at a voltage of 245 V, in agreement with previous measurements.
For the first time in a beam test, 25 $\mu$m (FBK25) and 35 $\mu$m (FBK35) thick FBK UFSDs sensors have been tested. 
The FBK35 sensor reaches a time resolution of 22 ps at a bias voltage of 240 V.
FBK25 reaches a time resolution of 25 ps at a bias voltage of 120 V. This result does not show an improvement with respect to FBK35, as the FBK25 sensors have been produced on a highly doped substrate (p-concentration $\sim$ 10$^{14}$ atoms/cm$^3$), not optimized for timing measurements; the low resistivity bulk increases the noise from the sensor and worsen the time resolution.
For all sensors, a good agreement between data and simulation has been obtained. The results  demonstrate the expectations of improved time resolution with thinner UFSD detectors, going from the standard 50 $\mu$m to a thinner design.

\backmatter

\bmhead{Acknowledgments}

We kindly acknowledge the following funding agencies and collaborations: INFN – FBK agreement on sensor production; Dipartimenti di Eccellenza, Univ. of Torino (ex L. 232/2016, art. 1, cc. 314, 337); Ministero della Ricerca, Italia, PRIN 2017, Grant 2017L2XKTJ – 4DinSiDe; Ministero della Ricerca, Italia, FARE,    Grant R165xr8frt\_fare.\\
The authors wish also to thank the support of the mechanical and electronic workshops of the INFN Unit of Bologna and the CERN-PS operator team for the support.

\section*{Declarations}
The study was funded by: INFN – FBK agreement on sensor production; Dipartimenti di Eccellenza, Univ. of Torino (ex L. 232/2016, art. 1, cc. 314, 337); Ministero della Ricerca, Italia, PRIN 2017, Grant 2017L2XKTJ – 4DinSiDe; Ministero della Ricerca, Italia, FARE, Grant R165xr8frt\_fare.\\
The authors received research support from institutes as specified in the author list beneath the title. 


\bibliography{sn-bibliography}


\end{document}